\documentclass[a4paper]{jpconf}
\usepackage{graphicx,amsmath,amssymb}

\newcommand{\beq}{\begin{equation}}
\newcommand{\eeq}{\end{equation}}
\newcommand{\bea}{\begin{eqnarray}}
\newcommand{\eea}{\end{eqnarray}}
\newcommand{\ba}{\begin{align}}
\newcommand{\ea}{\end{align}}
\newcommand{\bfig}{\begin{figure}}
\newcommand{\efig}{\end{figure}}

\newcommand{\D}{\displaystyle}

\newcommand{\gev}{\, \text{GeV}}

\newcommand{\thetaprime}{\theta^{\prime}}

\newcommand{\tin}{t_{\rm in}}
\newcommand{\la}{\langle}
\newcommand{\ra}{\rangle}

\newcommand{\tplus}{t_{+}}

\newcommand{\omnes}{{\cal{O}}}

\newcommand{\lprime}{\lambda^{\prime}}
\newcommand{\ldprime}{\lambda^{\prime \prime}}
\begin{document}

\title{Constraining Form Factors with the Method of Unitarity Bounds}

\author{B. Ananthanarayan$^{a,*}$
Irinel Caprini$^{b,\dag}$}

\address{$^a$Centre for High Energy Physics, Indian Institute of Science,
Bangalore 560 012, India}
\address{$^b$ Horia Hulubei National Institute for Physics and Nuclear
Engineering, P.O.B. MG-6, 077125, Maguerele, Romania}

\ead{$^*$anant@cts.iisc.ernet.in,\,
$^\dag$caprini@theory.nipne.ro}

\begin{abstract}
The availability of a reliable bound on an integral involving
the square of the modulus of a form factor  on the unitarity cut allows one to constrain 
the form factor at  points inside the analyticity domain and its shape parameters,
and also to isolate domains on the real axis and in 
the complex energy plane where zeros are excluded.  In this lecture note, we review the
mathematical techniques of this formalism in its standard form, known as the method of unitarity bounds, and 
recent developments which allow us to include information
on the phase  and modulus along a part
of the unitarity cut.  
We also provide a brief summary of some results that we have
obtained in the recent past, which demonstrate the usefulness of the method for precision predictions on the form factors. 
\end{abstract}

\section{Introduction}

Form factors are basic observables in strong interaction dynamics.
They provide information on the nature of the strong force and
confinement.  The pion electromagnetic form factor is one such
observable.  The weak form factors are of crucial importance for 
the determination of standard model parameters 
such as the elements of the Cabibbo-Kobayashi-Maskawa (CKM) matrix.

Before the advent of the modern theory of strong interactions,
mathematical methods were advanced to obtain
bounds on the form factors when a suitable integral of the modulus squared
along the unitarity cut $(t_+,\infty)$ was available \cite{PHRVA.D3.2807,PHRVA.D4.725}  
(for a topical review of the results at that time, see \cite{SiRa}). 
Using the methods of  complex analysis, this condition leads to constraints 
on the values at points inside the analyticity domain or on the derivatives  at $t=0$, 
 such as the slope  and curvature. Also of interest was whether or not the form factors can
vanish on the real energy or in the complex energy plane
(in the context of the pion electromagnetic form factor,
see, e.g. \cite{Raszillier:1976as}). Mathematically, these problems belong to a class 
referred to as the Meiman 
problem \cite{Meiman}.\footnote{Stated differently, the problem belongs 
to the standard analytic interpolation theory for functions in the 
Hardy class $H^2$.}

At present, effective theories of strong interactions, like Chiral Perturbation Theory (ChPT) or Heavy Quark Effective Theories (HQET), as well as lattice calculations or various types of QCD sum rules, allow us to make predictions, sometimes very precise, on the form factors at particular kinematical regions. However, a full calculation of the hadronic form factors from first principles is not yet possible.
On the other hand, experimental information on the form factors
has improved considerably in recent years. In these 
circumstances, analytic techniques as those mentioned above prove to be very useful,  providing rigorous correlations and consistency checks of various approaches and in improving  the phenomenological analyses of the form factors.
In a recent review, ref.~\cite{review} we have presented
a complete treament of the mathematical background required
for studying these problems and a comprehensive bibliography
on the subject.

The integral condition was provided either from an observable 
(like muon's $g-2$ in the case of the electromagnetic form factor of the pion)
or, in the modern approach first applied in \cite{NUPHA.B189.157},  from the dispersion relation satisfied by a suitable 
correlator calculated by perturbative QCD in the spacelike region, and
whose positive spectral function has, by unitarity, a lower bound 
involving the modulus squared of the relevant form factor.  
Therefore, the constraints derived in this framework are often 
referred to as ``unitarity bounds''.  
Information about the phase of the form factor, available
by Watson's theorem from the associated elastic scattering,
 can be used to improve the results
in a systematic fashion \cite{PHRVA.D7.2136,NUPHA.B98.204}.
In some cases the modulus is also measured independently  along a part of the unitarity cut.

Recent applications in the modern approach concerned mainly the form factors
relevant for  the $B\to D^{(*)}$ semileptonic decays, 
or the so-called Isgur-Wise function,  where heavy quark symmetry provided 
strong additional constraints at interior 
points~\cite{PHLTA.B282.215,hep-ph/9306214,hep-ph/9504235,hep-ph/9508211,hep-ph/9603414,hep-ph/9712417}. 
More recent applications revisited the electromagnetic form factor 
of the pion~\cite{Caprini2000,arXiv:0801.2023,arXiv:0811.0482,arXiv:0903.4297,pipi}, the strangeness changing $K\pi$ form factors~\cite{hep-ph/0504016,hep-ph/0607108,arXiv:0905.0951,pik1,pik2}, the $B\pi$ vector form factor \cite{hep-ph/9509358,hep-ph/0509090, arXiv:0807.2722}, and also
 the $D\pi$ form factors~\cite{dpi}. The results confirm that the approach 
represents a useful tool in the study of the form factors, complementary  
and free of additional assumptions inherent in standard dispersion relations.
The techniques presented here are the framework
of including inputs coming from disparate sources and 
effectively testing their consistency.
In this lecture note we will describe the mathematical machinery that has
been developed to address these issues.  
We will also review some recent results obtained by us using
these methods. 

 In Sec.~\ref{meiman} we  present our notation and describe the general Meiman 
interpolation problem, for an arbitrary number of derivatives at the origin and
an  arbitrary number of values at points inside the analyticity  
domain. 
The solution is obtained by Lagrange multipliers, 
and is written as a positivity condition of a determinant written down in terms of the input values.
Alternatively, using techniques of analytic interpolation theory, the result can be written as a compact convex quadratic form, given in  ref.~\cite{review}.  We present the
complete treatment of the inclusion of the phase on a part of the unitarity cut,
($t_+,\tin)$,
along with an arbitrary number of constraints 
of the Meiman type in Sec.~\ref{phase}.  The general treatment 
was only recently provided in entirety despite the 
lengthy literature on the problem.  Two equivalent sets of integral
equations have been found and here we provide the result obtained from using 
the method of  Lagrange multipliers.  The result obtained from  analytic interpolation theory can be found in \cite{review}. 
In Sec.~\ref{modulusphase}  we treat a modified analytic optimization problem,  
where the input on the cut consists from the phase below a certain threshold $\tin$ 
and a weighted integral over the modulus squared along $t>\tin$. 
This is a mathematically more complicated
problem solved for the first time  in \cite{Caprini2000}. The method
uses the fact that the knowledge of the phase  allows one to describe exactly
the elastic cut of the form factor by means of the Omn\`{e}s function.  The
problem is thus reduced to a standard Meiman problem  on a larger analyticity
domain. 
In practice, the integral  along $t>\tin$ can be obtained by subtracting from the 
integral along the whole cut the low-energy contribution, $t<\tin$, which can be estimated
 if some information on the modulus on that part of the cut is 
available from an independent source. The integral can be evaluated also if precise data on the modulus are available up to high energies, as is the case with the pion electromagnetic form factor \cite{BABAR,KLOE1}. 
In Sec.~\ref{applications} we present for illustration  most recent results
of these methods applied to 
the pion electromagnetic form factor~\cite{pipi}.
We conclude with an afterword, Sec.~\ref{afterword}.

\section{Meiman Problem}\label{meiman}

In order to set up the notation, let us begin by 
letting $F(t)$ denote a form factor, which is real  analytic ({\em i.e.} $F(t^*)=F^*(t)$)
in the complex $t$-plane with a cut along the positive real axis from the 
lowest unitarity branch point $\tplus$ to 
$\infty$. The essential condition considered in the present context 
is an inequality:
\beq
 \int^{\infty}_{\tplus } dt\ \rho(t) |F(t)|^{2} \leq I,
        \label{eq:I}
\eeq
where $\rho(t)\geq 0$ is a positive semi-definite weight function 
and $I$ is a known quantity. As mentioned in the Introduction, such
inequalities can be obtained starting from  a dispersion relation satisfied by a
suitable correlator,  evaluated  in the deep Euclidean region by perturbative 
QCD, and whose spectral function is  bounded from below by a term involving the
modulus squared of the relevant form factor.  

Of interest in the analysis of form factors are the shape parameters
that appear in the expansion around $t=0$.  For instance, 
in the case of the pion electromagnetic form factor 
the expansion is customarily written as
\begin{equation}
\label{eq:taylorpi}
	F(t) = 1 + \D\frac{1}{6} \la r^2_\pi \ra t + c t^2 + d t^3 + \cdots,
\end{equation}
whereas 
in the analysis of the semileptonic decays, one is interested in the 
shape parameters appearing in 
\beq\label{eq:taylor}
F(t)= F(0) \left[1+ \lprime\, \frac{t}{M^2}  + \ldprime\, \frac{t^2}{2
M^4} 
+ \cdots \right],
\eeq 
where $M$ is a suitable mass and  $\lprime$ and
$\ldprime$ denote the
dimensionless slope and curvature,
respectively.  

Low energy theorems, reflecting the
chiral symmetry of the strong interaction, and
lattice calculations can provide information on $F(t)$ at several special
points inside the analyticity domain.  The standard unitarity bounds exploit
analyticity of the form factor and the inequality (\ref{eq:I}) in order to
correlate in an optimal way these values and the expansion parameters in
(\ref{eq:taylor}).

In order to set up the stage,
the problem is brought to a canonical form by
making the conformal transformation
\beq\label{eq:z}
\bar z(t)=\frac{\sqrt{t_+}-\sqrt{t_+-t}}{\sqrt{t_+}+\sqrt{t_+-t}}\,,
\eeq
that maps the cut $t$-plane onto the unit disc $|z|<1$ in the $z\equiv \bar z(t)$ 
plane, such that $\tplus$ is mapped onto $z = 1$, the point at 
infinity to $z = -1$ and the origin to $z=0$. After this mapping, the
inequality 
(\ref{eq:I}) is written as
\beq\label{eq:gI}
\frac{1}{2 \pi} \int_{0}^{2\pi} {\rm d}\theta |g(e^{i\theta})|^2 
	\leq I,
\eeq
where the analytic function $g(z)$ is defined as
\beq\label{eq:gz}
g(z) = F(\bar t(z)) w(z).	
\eeq
Here  $\bar t(z)$ is the inverse of (\ref{eq:z}) and  $w(z)$ is an 
{\it outer function}, {\it i.e.}  a function analytic and without zeros in
$|z|<1$, such that its modulus on the boundary is related to $\rho(\bar t(\e^{i\theta}))$
and the Jacobian of the transformation (\ref{eq:z}). In particular cases of physical interest, 
the outer functions $w(z)$ have a simple analytic 
form. In general, an outer
function is obtained from its modulus on the boundary by the integral
\beq\label{eq:w}
w(z)=\exp\left[\frac{1}{2\pi} \int_{0}^{2\pi} {\rm d}\theta \,
\frac{e^{i\theta}+z}{e^{i\theta}-z}\,\ln |w(e^{i\theta})| \right].
\eeq

The function $g(z)$ is analytic within 
the unit disc and can be expanded as:
\beq\label{eq:gTaylor}	
g(z)=g_0+g_1 z+ g_2 z^2 + \cdots,
\eeq
and (\ref{eq:gI}) implies
\beq\label{eq:gkI}
\sum_{k=0}^\infty g_k^2 \leq I.
\eeq
Using (\ref{eq:gz}), the real numbers $g_k$  are expressed in a straightforward
way  in terms of the coefficients of the Taylor expansion (\ref{eq:taylor}). The
inequality (\ref{eq:gkI}), with the sum in the left side truncated at some finite order,
 represents the simplest ``unitarity bound" for the
shape parameters defined in (\ref{eq:taylor}). In what follows we shall improve
it by including additional information on the form factor. 

We consider the general case when the first $K$ derivatives of $g(z)$ at $z=0$
and the values at $N$ interior points  are assumed to be known:
\bea\label{eq:cond}
\left[\D \frac{1}{k!} \D \frac{ d^{k}g(z)}{dz^k}\right]_{z=0}&=& g_k, \quad
0\leq k\leq K-1; \nonumber\\
 g(z_n)&=&\xi_n, \quad  1\leq n \leq N, 
\eea
where $g_k$ and $\xi_n$ are given numbers.  They  are related, by (\ref{eq:gz}),
to the derivatives $F^{(j)}(0)$, $j\le k$  of $F(t)$ at $t=0$, and the values
$F(\bar t(z_n))$, respectively. For simplicity
and in view of phenomenological inputs that we will use,
we assume the points $z_n$ to be real, so $\xi_n$ are also real. 
The Meiman problem  requires to find the optimal
constraints satisfied by the 
numbers defined in (\ref{eq:cond})  if (\ref{eq:gI}) holds.  
One can prove that the most general constraint 
satisfied by the input values appearing in (\ref{eq:cond}) is given by the
inequality:
\beq\label{eq:domain}
\mu_0^2 \leq I,
\eeq
where $\mu_0^2$ is the solution of the minimization problem:
\beq\label{eq:min}
\mu_0^2= \min\limits_{g\in {\cal G}} \, ||g||^2_{L^2}.
\eeq
Here $ ||g||_{L^2}^2$ denotes the $L^2$ norm, {\it i.e.} the quantity appearing 
in the l.h.s. of (\ref{eq:gI}) or (\ref{eq:gkI}), and the minimum is taken over
the class   ${\cal G}$ of analytic functions  
which satisfy  the conditions (\ref{eq:cond}).

The minimization problem (\ref{eq:min}) can be solved
by using the Lagrange multiplier method. The Lagrangian may be written as
\beq\label{eq:L}
{\cal L} = 
\frac{1}{2} \sum_{k = 0}^{\infty} g_k^2 + \sum_{n=1}^N
\alpha_n (\xi_n - \sum_{k = 0}^{\infty} g_k z^k),
\eeq
where $\alpha_n$  are real Lagrange multipliers.
Solving the Lagrange equations obtained by varying with respect 
to $g_k$ for all $k \ge  K$, and eliminating the Lagrange  multipliers yields
the solution of the minimization problem (\ref{eq:min}).
For purposes of illustration, when $N = 2$, the Lagrange equations yield
\beq
 \displaystyle g_k=\alpha_1 z_1^k +\alpha_2 z_2^k, \quad \quad k\ge K,
	\label{eq:lagn2}
\eeq
and the  inequality (\ref{eq:domain}) can be expressed in terms of the two Lagrange
multipliers as:
\beq
 \displaystyle 
\alpha_1 \bar{\xi}_1 + \alpha_2 \bar{\xi}_2 \,\leq \,\bar{I},
\label{eq:constn2}
\eeq
where  $\bar{\xi}_n$ are known numbers defined as 
\beq
\bar{\xi}_n = \xi_n - \sum_{k=0}^{K-1}g_k z_n^k,
\eeq
and 
\beq\label{eq:tildeI}
\bar{I} = I - \sum_{k = 0}^{K-1} g_k^2.
\eeq
 The constraint conditions themselves are
\bea\label{eq:xi1xi2}
& \displaystyle 
\alpha_1 \frac{z_1^{2 K}}{1-z_1^2} + \alpha_2
\frac{(z_1 z_2)^{K}}{1-z_1 z_2} =\bar{\xi}_1,  \nonumber \\
& \displaystyle 
\alpha_1 \frac{(z_1 z_2)^K}{1-z_1 z_2}+
\alpha_2 \frac{z_2^{2 K}}{1-z_2^2}= \bar{\xi}_2.
\eea
The consistency of eqs. (\ref{eq:constn2}) and (\ref{eq:xi1xi2}) as a system of equations for 
$\alpha_i$ can be written
as:
\beq
\left|
\begin{array}{c c c}
\bar{I} & \bar{\xi}_1 & \bar{\xi}_2 \\
\bar{\xi}_1 & \D \frac{z_1^{2 K}}{1-z_1^2} & \D \frac{(z_1 z_2)^K}{1-z_1z_2} \\
\bar{\xi}_2 & \D \frac{(z_1 z_2)^K}{1-z_1 z_2} & \D \frac{z_2^{2 K}}{1-z_2^2}\\
\end{array}
\right| \geq 0.
\eeq
This can be readily extended to the case of $N$ constraints: 
\beq\label{eq:determinant}
\left|
	\begin{array}{c c c c c c}
	\bar{I} & \bar{\xi}_{1} & \bar{\xi}_{2} & \cdots & \bar{\xi}_{N}\\	
	\bar{\xi}_{1} & \D \frac{z^{2K}_{1}}{1-z^{2}_1} & \D
\frac{(z_1z_2)^K}{1-z_1z_2} & \cdots & \D \frac{(z_1z_N)^K}{1-z_1z_N} \\
	\bar{\xi}_{2} & \D \frac{(z_1 z_2)^{K}}{1-z_1 z_2} & 
\D \frac{(z_2)^{2K}}{1-z_2^2} &  \cdots & \D \frac{(z_2z_N)^K}{1-z_2z_N} \\
	\vdots & \vdots & \vdots & \vdots &  \vdots \\
	\bar{\xi}_N & \D \frac{(z_1 z_N)^K}{1-z_1 z_N} & 
\D \frac{(z_2 z_N)^K}{1-z_2 z_N} & \cdots & \D \frac{z_N^{2K}}{1-z_N^2} \\
	\end{array}\right| \ge 0.
\eeq
Alternatively, the solution
can be obtained by introducing  Lagrange multipliers also for the given
coefficients
$g_k,\, k=0,...,K-1$ in (\ref{eq:L}). 
This leads to an inequality, equivalent to (\ref{eq:determinant}), written as \cite{SiRa2}:
\beq
\hspace*{-0.2in}\left|
\begin{array}{c c c  c c c c c c}
I & g_0 & g_1 & \cdots & g_{K-1}&  \xi_1 & \xi_2 & \cdots &\xi_N \\
g_0 & 1 & 0 & 0 & 0   & 1 & 1 & \cdots & 1\\
g_1 & 0 & 1 & 0 & 0 & z_1 & z_2 & \cdots & z_N \\
\vdots & \vdots & \vdots  & \vdots & \vdots & \vdots & \vdots & \vdots
& \\
g_{K-1} & 0 & 0 &  \cdots & 1 & z_1^{K-1} & z_2^{K-1} & \cdots & z_N^{K-1}
\\
\xi_1 & 1 & z_1 &  \cdots & z_1^{K-1}& \frac{1}{1-z_1^2} & \frac{1}{1-z_1 z_2}
& 
\cdots & \frac{1}{1-z_1 z_N}\\
\xi_2 & 1 & z_2  & \cdots & z_2^{K-1}& \frac{1}{1-z_1z_2} & \frac{1}{1- z_2^2}
& 
\cdots & \frac{1}{1-z_2 z_N}\\
\vdots & \vdots  & \vdots & \vdots&\vdots & \vdots & \vdots & \vdots &\\
\xi_N & 1 & z_N &  \cdots & z_N^{K-1}& \frac{1}{1-z_1z_N} & \frac{1}{1- z_2 z_N}
& 
\cdots & \frac{1}{1-z_N^2}\\
\end{array}\right| \ge 0. 
\label{eq:det1}
\eeq
The  conditions (\ref{eq:determinant}) or  (\ref{eq:det1}) can be expressed in a straightforward way in terms of the values of
the form factor $F(t)$ at  $t_i=\bar t(z_i)$ and the derivatives at $t=0$, using eqs.
(\ref{eq:z}) and (\ref{eq:gz}). It can be shown that these inequalities define 
convex domains in the space of the input parameters.  
\section{Inclusion of Phase Information}\label{phase}
Additional information on the unitarity cut can be included in the formalism.
According to the Fermi-Watson theorem, below the  
inelastic threshold $\tin$ the phase of $F(t)$  is equal (modulo $\pi$) to the 
phase $\delta(t)$ of the associated elastic scattering process. Thus,
\beq\label{eq:watson}
F(t+i\epsilon)= |F(t)| e^{i\delta(t)}, \quad \quad t_+<t< \tin,
\eeq
where $\delta(t)$ is known. 

In order to impose the constraint  (\ref{eq:watson}), we  define first the Omn\`{e}s function
\beq	\label{eq:omnes}
 \omnes(t) = \exp \left(\D\frac {t} {\pi} \int^{\infty}_{\tplus} dt 
\D\frac{\delta (t^\prime)} {t^\prime (t^\prime -t)}\right),
\eeq
where $\delta(t)$  is  known for 
$t\le \tin$, and is an arbitrary function, sufficiently  smooth ({\em i.e.}
Lipschitz continuous) for $t>\tin$. From  (\ref{eq:watson}) and (\ref{eq:omnes})
it follows that
\beq\label{eq:ratio}
{\rm Im} \left[\frac{ F(t + i\epsilon)}{ \omnes(t +i\epsilon )}\right]=0, 
\quad\quad  t_+\le t \le \tin.
\eeq
 Expressed in terms of the function $g(z)$ this condition becomes
\beq\label{eq:img}
{\rm Im} \left[\frac{g (e^{i\theta})} {W(\theta)}\right] =0, \quad \quad 
\theta \in (-\theta_{\rm in},\theta_{\rm in}).
\eeq
Here $\theta_{\rm in}$ is defined by $\bar z(\tin)=\exp(i\theta_{\rm in})$ and  
the function $W(\theta)$ is defined as:
\beq\label{eq:W}
W(\theta)= w(e^{i\theta} ) O(e^{i\theta}),
\eeq 
where  $w(z)$  is the outer function and 
\beq\label{eq:O}
O(z)=\omnes(\bar t(z)). 
\eeq
The constraint (\ref{eq:img}) can be 
imposed by means of a generalized Lagrange multiplier, while
constraints at interior points can be treated with standard real-valued Lagrange
multipliers.  The Lagrangian of the minimization problem (\ref{eq:min}) 
with the constraints (\ref{eq:cond}) and (\ref{eq:img}) reads
\bea
{\cal L} &=& \frac{1}{2} \sum_{k = 0}^{\infty} g_k^2 + \sum_{n=1}^N
\alpha_n (\xi_n - \sum_{k = 0}^{\infty} g_k z^k)\\ 
&+& \frac{1}{\pi} \sum_{k = 0}^{\infty} g_k \lim_{\rm r \rightarrow 1} 
\int\limits_{-\theta_{\rm in}}^{\theta_{\rm in}} \lambda(\theta') 
|W(\theta')| {\rm {Im}} [[W(\theta')]^{-1} r^k e^{i k \theta'}] 
d\theta'.\nonumber 
\eea
The Lagrange multiplier $\lambda(\theta)$ is an odd function, 
$\lambda(-\theta) = -\lambda(\theta)$ and                    
the factor  $|W(\theta')|$ was introduced in the integrand for convenience.

We  minimize ${\cal L}$ by brute force method with respect to the free 
parameters $g_k$ with $k\geq K$. The Lagrange multipliers $\lambda(\theta)$ and
$\alpha_n$  are  found in the standard way by imposing the
constraints (\ref{eq:cond}) and (\ref{eq:img}).

The calculations are straightforward.  In order 
to write the equations in a simple form, it is convenient to define the phase
$\Phi(\theta)$ of the function $W(\theta)$ by
\beq
W(\theta) = |W(\theta)| e^{i \Phi(\theta)}.
\label{wtheta_eqn1}
\eeq
From (\ref{eq:W}) we have
\beq
\Phi(\theta) = \phi(\theta) + \delta (\tilde t(e^{i\theta})),
\label{phi_eqn1}
\eeq 
where $\phi(\theta)$ is the phase of the outer function $w(e^{i\theta})$ 
and $\delta(t)$ is the  elastic scattering phase shift. We introduce also the
functions $\beta_n$ for $n=1,...N$, by
\beq\label{eq:betan}
\beta_n(\theta)=z_n^{K} \,\frac{\sin[K \theta-\Phi(\theta)]-z_n 
\sin[(K-1)\theta-\Phi(\theta)]}{1+z_n^2-2 z_n\cos\theta}.
\eeq
Then the  equations for the Lagrange multipliers $\lambda(\theta)$ and $\alpha_n$ 
take the form:
\bea\label{eq:eq1}
&& \sum_{k = 0}^{K-1} g_k \sin[k \theta - \Phi(\theta)] =\lambda(\theta) 
-\sum_{n=1}^N \alpha_n \beta_n(\theta)   \\                     
&&\hspace{-0.1cm} - \frac{1}{2\pi} \int\limits_{-\theta_{\rm in}}^{\theta_{\rm
in}} 
{\rm d} \thetaprime \lambda(\thetaprime) ) {\cal K}_{\Phi}(\theta, \thetaprime),
\quad \quad \theta \in (-\theta_{\rm in},\theta_{\rm in}),  \nonumber 
\eea
\beq\label{eq:eq2}
- \frac{1}{\pi} \int\limits_{-\theta_{\rm in}}^{\theta_{\rm in}} 
\lambda(\theta)\beta_n(\theta) {\rm d}\theta 
+ \sum_{n'=1}^N \alpha_{n'}  \frac{(z_n z_{n'})^{K}}{1-z_nz_{n'}} = \bar{\xi}_n, 
\eeq
where  $n=1,\ldots N$.  

The integral kernel in (\ref{eq:eq1}), defined as
 \beq\label{eq:calK}
{\cal K}_{\Phi}(\theta, \thetaprime) \equiv \frac{\sin[(K-1/2) (\theta -
\thetaprime) - \Phi(\theta) +
\Phi(\theta^\prime)]}{\sin[(\theta-\theta^\prime)/2]}, 
\eeq
is of Fredholm type if the phase $\Phi(\theta)$ is Lipschitz continuous 
Then the  above system can be solved numerically in a straightforward manner. 
Finally, the inequality (\ref{eq:domain}) takes the form:
\beq
\label{eq:domain2}
\frac{1}{\pi} \sum_{k = 0}^{K-1} g_k \int\limits_{-\theta_{\rm
in}}^{\theta_{\rm
in}} 
{\rm d} \theta \lambda(\theta) 
\sin\left[k \theta - \Phi(\theta) \right] +\sum_{n=1}^N \alpha_n \bar{\xi}_n \leq \bar{I},
\eeq
with $\bar I$ defined in (\ref{eq:tildeI}). Using the relation (\ref{eq:gz}), 
the above inequality defines an allowed domain for the values of the form factor and its derivatives at the origin. 
It must be emphasized that the theory for arbitrary number of constraints 
was presented for the first time in ref.~\cite{review}.
It is easy to see that, if $\tin$ is increased, the allowed domain 
defined by the 
inequality  (\ref{eq:domain2}) becomes smaller. 
The reason is that by increasing $\tin$  the class of functions 
entering the minimization (\ref{eq:min}) becomes 
gradually smaller, leading to a larger value for minimum $\mu_0^2$ entering the 
definition (\ref{eq:domain}) of the allowed domain.

\section{A Modified Optimization Problem}\label{modulusphase}
In this section we shall present the solution of a modified interpolation problem, 
where the input consists from the phase condition (\ref{eq:watson}) and the inequality
\beq\label{eq:Iprime}
\int_{\tin}^\infty {\rm d}t \rho(t)|F(t)|^2 \leq I^\prime,
\eeq
where $I^\prime$  is known. Unlike the previous condition (\ref{eq:I}), which involved the whole unitarity cut, 
we assume now that an integral of the modulus squared along $t>\tin$ (or a reliable upper bound on it) is known. 

In some cases, the  quantity $I^\prime$ can be obtained from the inequality (\ref{eq:I}) and the modulus of the form factor along the elastic part 
of the unitarity cut,  if the latter is available from an independent source, for instance from experiment. Then  $I^\prime$ is given by
\beq\label{eq:I1}
I^\prime= I - \int^{\tin}_{\tplus} {\rm d}t \rho(t) |F(t)|^2 <I.
\eeq
In other cases, like for the pion electromagnetic form factor,  $I^\prime$ can be estimated directly from experimental data on the modulus at high energies, taking into account also the asymptotic decrease  $|F(t)|\sim 1/t$ predicted by perturbative QCD. 

Once the conditions  (\ref{eq:watson}) and (\ref{eq:Iprime}) are adopted, one can find the optimal domain allowed for
the values and the derivatives of the form factor defined in (\ref{eq:cond}). Below we present the solution of the problem as first described in \cite{Caprini2000}. 
We start with  the remark that the knowledge of the phase was
implemented in the previous section by the relation 
(\ref{eq:ratio}), which says that the function $f(t)$ defined through
\beq
F(t) = f(t) \omnes(t),
\eeq
is real in the elastic region, $\tplus \le t \le \tin$. In fact, since the Omn\`{e}s function 
$\omnes(t)$ 
fully accounts for  the elastic cut of the form factor,  the function $f(t)$ has
a larger analyticity domain, namely the complex $t$-plane cut only for $t>\tin$.
 Moreover, (\ref{eq:I1}) implies that $f$ satisfies the  condition 
\beq\label{eq:fI}
\int_{\tin}^\infty {\rm d}t \rho(t) |\omnes(t)|^2 |f(t)|^2 \leq I^\prime.
\eeq
 This inequality leads, through the techniques presented in section \ref{meiman}, 
to constraints on the values of $f$  inside the analyticity domain. It is easy to see that the problem differs from the standard one described there only by the appearance of the additional  factor $|\omnes(t)|^2$ in the integral, and the fact that the cut starts now at $\tin>\tplus$.

The problem is brought into a canonical form by the new transformation
\beq\label{eq:ztin}
\tilde z(t) = \frac{\sqrt{\tin}-\sqrt {\tin -t} } {\sqrt {\tin}+\sqrt {\tin -t}}\,,
\eeq
which maps the complex $t$-plane cut for $t>\tin$ on the unit disc in the $z$-plane defined by $z=\tilde z(t)$. Then
(\ref{eq:fI}) 
can be written as
\beq\label{eq:gI1}
\frac{1}{2 \pi} \int^{2\pi}_{0} {\rm d} \theta |g(\exp(i \theta))|^2 \leq I^\prime,
\eeq
where  the function $g$ is now
\beq\label{eq:gtilde}
 g(z) = w(z)\, \omega(z) \,F(\tilde t(z)) \,[O(z)]^{-1}.
\eeq 
Here  $w(z)$ is the outer function related to the weight $\rho(t)$ and 
the Jacobian of the new mapping (\ref{eq:ztin}) and $O(z)$ is defined as
\beq\label{eq:Otilde}
O(z) = \omnes(\tilde t(z)),
\eeq
where  $\tilde t(z)$ is the inverse of $z=\tilde z(t)$ with $\tilde z(t)$ defined in (\ref{eq:ztin}), and
\beq\label{eq:omegatin}
 \omega(z) =  \exp \left(\D\frac {\sqrt {\tin - \tilde t(z)}} {\pi} \int^{\infty}_{\tin} {\rm d}t^\prime \D\frac {\ln |\omnes(t^\prime)|}
 {\sqrt {t^\prime - \tin} (t^\prime -\tilde t(z))} \right).
\eeq 
The inequality (\ref{eq:gI1}) has exactly the same form as (\ref{eq:gI}), and leads to the constraints (\ref{eq:determinant}) or (\ref{eq:det1}) for the values and derivatives of $g(z)$ at interior points.
Using (\ref{eq:gtilde}), these
constraints are expressed in terms of the physically interesting values of the
form factor $F(t)$. 

 A remark on the uniqueness is of interest: we recall that
the Omn\`es function $\omnes(t)$ defined in (\ref{eq:omnes}) is not unique, 
as it
involves the arbitrary function $\delta(t)$ for $t>\tin$.  In section \ref{phase} we have seen
that the results are not affected by this
arbitrariness, as the integral equations involved only the known phase below $\tin$. 
This is true also for the results here: the reason is that a change
of the function  $\delta(t)$ for $t>\tin$ is equivalent with a multiplication of
$g(z)$ by a function analytic and without zeros  in $|z|<1$ ({\it i.e.} an outer
function). According to the general theory of analytic functions of Hardy class, the multiplication by an outer function does not change the class of 
functions used
in minimization problems. In our case,  the arbitrary function
$\delta(t)$ for $t>\tin$ enters in both the functions $O(z)$ and  $\omega(z)$
appearing in (\ref{eq:gtilde}), and their ambiguities compensate each other
exactly. 
The independence of the results on the choice  of the phase  for $t>\tin$ is
confirmed  numerically, for functions $\delta(t)$ that are Lipschitz
continuous.
The constraints provided by the technique of this section are expected to be
quite strong since they result from a minimization on a restricted class of
analytic functions, where the second Riemann sheet of the form factor is
accounted for explicitly  by the Omn\`es function. 

On the other hand, it is easy to
see that the fulfillment of the condition (\ref{eq:fI}) does not automatically
imply that the original condition (\ref{eq:I}) is satisfied, 
and both conditions should be applied in order to reduce 
the allowed range of the parameters of interest.

The mathematical techniques presented can be adapted in a straightforward way  
to  the problem of zeros.  Let us assume that 
the form factor $F(t)$ has a simple zero on the real axis, $F(t_0)=0$. 
We shall use this information in the 
determinant condition: if the zero 
is compatible with the remaining information, the inequality 
can be satisfied. If, on the contrary, the inequality 
is violated, the zero is excluded. It follows that we can obtain 
a rigorous condition for the domain of points 
$z_0$ (or $t_0$) where the zeros are excluded. 

For illustration, assume first that we use as input only the value of the form factor at $t=0$. 
Then from the argument given above,
it follows that the domain of real points $t_0$ where the form factor cannot have zeros is described by the inequality \cite{pipi}
\beq	\label{eq:detz1}
\left|
	\begin{array}{c c }
	I'-g_{0}^{2} & -g_0\\	
	-g_0 & \frac{\tilde z(t_0)^2}{1-\tilde z(t_0)^2}  \\
	\end{array}\right|\leq 0.
\eeq
Here, $I^\prime$ is the known bound appearing in (\ref{eq:Iprime}), $g_0=g(0)$ is calculated in terms of the input 
phase from (\ref{eq:gtilde}),
and $\tilde z(t)$ is defined in (\ref{eq:ztin}). 
If we include in addition the value $F(t_1)$ of the form factor
at some point $t_1$, the condition reads \cite{pipi}:

\beq	\label{eq:detz2}
\left|
	\begin{array}{c c c }
	I'-g_{0}^{2} & g(\tilde z(t_1))-g_0 & -g_0\\	
	g(\tilde z(t_1))-g_0 &  \frac{\tilde z(t_1)^2}{1-\tilde z(t_1)^2} & \frac{\tilde z(t_1)\tilde z(t_0)}
  {1-\tilde z(t_1)\tilde z(t_0)}\\
	-g_0 &\frac{\tilde z(t_1)\tilde z(t_0)}{1-\tilde z(t_1)z(t_0)} & \frac{\tilde z(t_0)^2}{1-\tilde z(t_0)^2}  \\
	\end{array}\right|\leq 0.
\eeq

\section{Recent Applications to the Pion Electromagnetic Form Factor}
\label{applications}
We will now describe some applications of the methods discussed above.
Whilst there has been a long series of applications of these methods
to the pion electromagnetic form factor, ref.~\cite{pipi} captures
the most stringent results.  This  is based on the fact
that the phase of the form factor is determined up to the
$\pi\omega$ threshold  ($\sqrt{\tin}$=917 MeV) in terms of the $I=1,\, l=1$
partial wave of $\pi\pi$ elastic scattering. We have used 
the recent parametrization given in ref.~\cite{PHRVA.D77.054015},
which agrees well 
with the solutions of the
Roy equations~\cite{hep-ph/0005297} between threshold and 800 MeV. 
The modulus $|F(t)|$ has been measured also recently with improved precision by 
BaBar \cite{BABAR} and KLOE \cite{KLOE1} collaborations.

We consider the two-pion contribution to the muon $g-2$, when
the weight $\rho(t)$ has the form
\bea\label{eq:amu}
\rho(t) = \frac{\alpha^2 M_{\mu}^2}{12 \pi} \frac{(t - t_+)^{3/2}}{t^{7/2}} K(t), \nonumber\\
K(t) = \int_0^1 du\, \frac{(1-u) u^2} {1- u + M_\mu^2 u^2/t}.
\eea
The two-pion contribution to muon anomaly was evaluated recently with great precision 
\cite{arXiv:0908.4300} from the accurate BaBar data on the modulus \cite{BABAR}. 
In ref. \cite{pipi} the value of the integral $I'$ defined in (\ref{eq:Iprime}) for the weight 
(\ref{eq:amu}) and the above choice of $\sqrt{\tin}$ was
evaluated to be $22.17 \times 10^{-10}$. 

A concrete application of these methods is one of constraining the
shape parameters appearing in (\ref{eq:taylorpi})
using as input the conditions (\ref{eq:watson}) and (\ref{eq:Iprime}) and
 the technique  described in section \ref{modulusphase}. We use as input also
the precise estimate of the coefficient of
the linear term  $\la r^2_\pi \ra = (0.435\pm 0.005)\, {\rm fm}^2$ 
 given in \cite{Colangelo}, and  additional spacelike data coming from  
\cite{Horn,Huber}, which are given in
Table  \ref{table:Huber}, where the first error is statistical and 
the second is systematic. 

\begin{table}
\begin{center}
\caption{Spacelike data from \cite{Horn,Huber}.}
\label{table:Huber}
\renewcommand{\tabcolsep}{1.5pc} 
\renewcommand{\arraystretch}{1.1} 
\begin{tabular}{c c c}
\hline\noalign{\smallskip}
 $t$&  Value[$\gev^2$] & $F(t)$  \\
\noalign{\smallskip}\hline\noalign{\smallskip}    	
$t_1$ & $-1.60$ & $0.243 \pm  0.012_{-0.008}^{+0.019}$  \\ 
$t_2$ &$ -2.45 $ & $ 0.167 \pm 0.010_{-0.007}^{+0.013}$  \\
\noalign{\smallskip}\hline
\end{tabular}
\end{center}
\end{table}

\begin{figure}[htb]
  	\begin{center}
\vspace{0.35cm}
  \includegraphics[width = 6.cm]{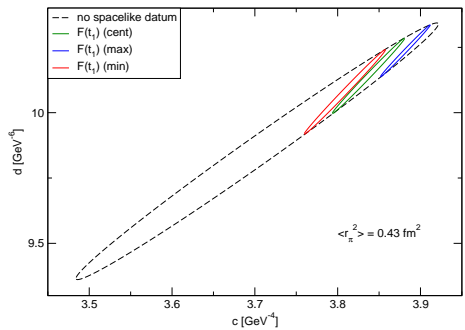}
\caption{Allowed domain in the $c-d$ plane calculated with  $\tin = (0.917 \, \gev)^2$ and $\la r^2_\pi \ra = 0.43 \,{\rm fm}^2$,  
for three values   of $F(t_1)$ at the spacelike point $t_1=-1.6\,\gev^2$ (central value in  Table \ref{table:Huber} 
and the extreme values obtained from the error intervals). Also shown is the large ellipse when no spacelike datum is included.}
\label{fig:figcd}
 	\end{center}
\end{figure}

An illustrative example of the results on the shape parameters $c$ and $d$ is given in Fig.~\ref{fig:figcd}.
Varying  $F(t_1)$ given in Table \ref{table:Huber} inside the error bars, we obtain the allowed domain of the 
$c$ and $d$ parameters  at the present level of knowledge as the union 
of the three  small  ellipses in Fig.  \ref{fig:figcd}.
By varying also $\la r^2_\pi \ra$ in the range given above the allowed ranges read 
\bea\label{eq:cdnum2}
3.75\,\gev^{-4} \lesssim c \lesssim 3.98\,\gev^{-4},\nonumber
\\  9.91\,\gev^{-6}  \lesssim d \lesssim 10.45\,\gev^{-6}, 
\eea
with a strong correlation between the two coefficients. 
Similar results are
obtained when the second datum in Table \ref{table:Huber} is used in place of the first.

Turning now to the issue of zeros, and using the machinery as described
in the preceding section, we find that, for
with $\la r^2_\pi \ra = 0.43\,{\rm fm}^2$, 
simple zeros are excluded from the interval  
$-1.93\,\gev^2 \leq t \leq 0.83\,\gev^2$ 
of the real axis.  If we  impose the additional constraint at 
a spacelike point $t_1=-1.6\,\gev^2$,  
the interval for the excluded zeros is much bigger. 
The left end of the range is 
sensitive to the input value $F_1=F(t_1)$.  
Using  the central value $F_1=0.243$ given in 
Table \ref{table:Huber}, we find 
that the form factor cannot  have simple 
zeros in the range $-5.56\, \gev^2 \leq t \leq 0.84\, \gev^2$.  
By varying $F_1$ inside the 
error interval given in Table \ref{table:Huber} 
(with errors added in quadrature), we find that 
zeros are excluded from the range 
$-4.46\, \gev^2 \leq t \leq 0.84\, \gev^2$.

\begin{figure}[htb]
  	\begin{center}
\vspace{0.35cm}
 \includegraphics[width = 5.6cm]{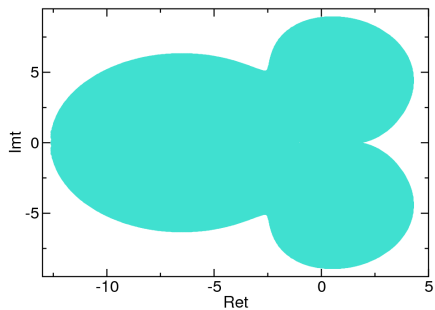}
\caption{Domain without zeros obtained  with $\tin = (0.917\, \gev)^2$ and  $\la r^2_\pi \ra = 0.43\, {\rm fm}^2$, 
using in addition   the central experimental value  $F(t_1)=0.243$ at the spacelike point $t_1=-1.6\,\gev^2$.}
	\label{fig:figz1}
 	\end{center}
\end{figure}

The machinery may now be used to find regions in the complex
plane where no zeros are allowed. Complex zeros occur  simultaneously at
 complex conjugate points due to the reality condition.
 In Fig.~\ref{fig:figz1}
we present the exclusion region obtained with central values
of the radius and the first datum in Table\ref{table:Huber}.
As in the case of the simple zeros on the real axis, there is significant
sensitivity to the experimental uncertainty at that point, the
allowed region being obtained as the intersection of the
regions corresponding to the extreme values.  The net effect is a somewhat reduced region
compared to that given in Fig.~\ref{fig:figz1}.
A detailed discussion on the sensitivity is given in ref.~\cite{pipi}.

\section{Afterword}\label{afterword}

In this lecture note we have provided
a comprehensive introduction to the modern theory of unitarity
bounds for hadronic form factors. 
As an illustration we have shown how  these methods can be exploited
 to improve the knowledge on the electromagnetic
form factor of the pion at low energies.  In \cite{arXiv:0905.0951,pik1,pik2} the techniques
were applied to the
pion-kaon form factors, which represent an important input for the
extraction of the CKM matrix element $V_{us}$,  while in \cite{dpi} 
the techniques were applied
the $D\pi$ form factors, of interest for the extraction
of the CKM matrix element $V_{cd}$.  The results of these
investigations are presented 
 in the contributions \cite{Kpi} and \cite{Dpi} to these Proceedings, respectively.
An important conclusion of these analyses is 
that there is a remarkable coherence between various
theoretical approaches to the strong interactions, including perturbative
QCD, chiral symmetries, lattice evaluations, and also experimental
information which is now available at high precision.

\section*{Acknowledgments}
We wish to thank the organizers of the workshop for inviting us
to present our work.
We thank Gauhar Abbas, I. Sentitemsu Imsong
and Sunethra Ramanan for discussions.

\end{document}